\documentclass[english,a4paper,12pt]{article}
\usepackage{amssymb,graphicx,hyperref}

\newcommand{\be}{\begin{equation}}
\newcommand{\ee}{\end{equation}}

\begin{document}

\title{From the cosmological model to generation of \\ the Hubble flow}

\author{V.\,N. Lukash, E.\,V. Mikheeva\thanks{E-mail: helen@asc.rssi.ru}\,, V.\,N. Strokov\thanks{E-mail: strokov@asc.rssi.ru}\\
         Astrospace Centre of the Lebedev Physical Institute}
\maketitle

\begin{abstract}
We review different approaches to the problem of generating the
observed Hubble flow, whose discussion was pioneered by
A.D.~Sakharov. Extrapolating the Cosmological Standard Model to
the past makes it possible to determine physical properties of and
conditions in the early universe. We discuss a new cosmogenesis
paradigm based on studying geodesically complete geometries of
black/white holes with integrable singularities.
\end{abstract}

\section{Introduction}

In his works (e.g.~\cite{sah-multiverse, sah-signature})
A.D.~Sakharov repeatedly mentioned the idea that cosmological
flows can be generated from ultra-high density states of matter
through quantum transitions resulting in different values of the
world constants, signature, time arrow and other geometrical
features of space--time. The way how gravitational systems and
parts thereof enter and leave such critical states remains a point
of controversy up till now.

High curvature and density are naturally generated in separate
spacetime domains in the course of collapse of compact
astrophysical objects that end up as black holes. However, the
problem of how collapse turns into expansion is yet to be solved.
The paradigm of multi-sheet universe proposed by A.D.~Sakharov as
well as the widely accepted Multiverse paradigm (see, for
example,~\cite{carr}) requires a clear and straightforward
physical mechanism of generating multiple expanding matter flows.
Still, the mechanism of {\it cosmogenesis} is rather vague. In
order to study it we need to know the structure of the initial
material flow called the early universe with sufficient precision.

\begin{figure}[t!]
\begin{center}
{\includegraphics{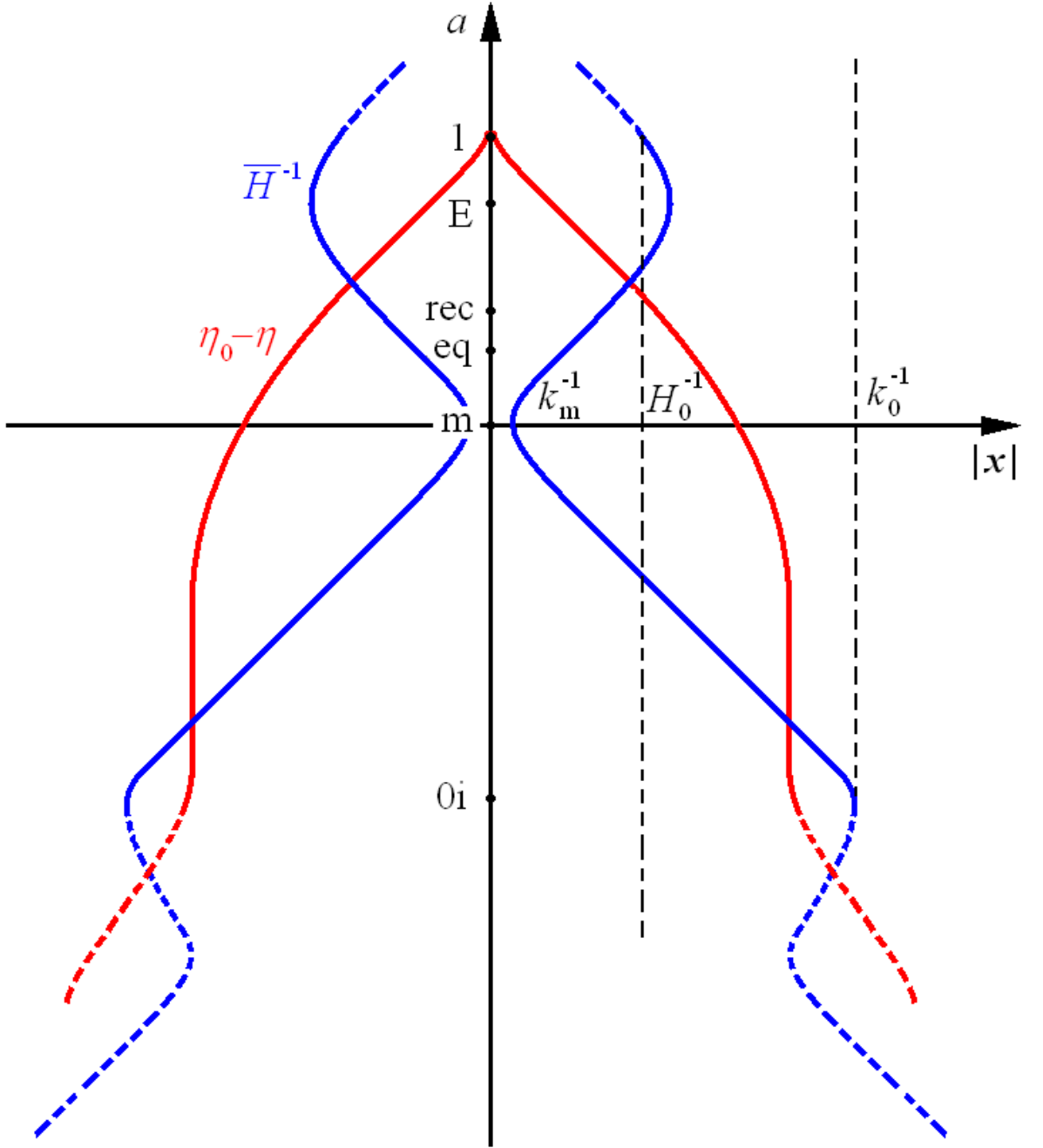}}
\end{center}
\caption{Vertical axis shows the scale factor $a$ while the
horizontal one -- the comoving coordinate $|\mathbf{x}|$ (an
observer's world line is $\mathbf{x}=0$). The line
$\overline{H}^{\,-1}$ corresponds to the Hubble
radius($\overline{H}=aH$), and $(\eta_{0}-\eta)$ is the light cone
of the past. The scales $k_{\rm{m}}^{-1}$, $H_{0}^{-1}$, and
$k_{0}^{-1}$, correspond to submillimeters, 4.3~Gpc, and the size
of the Friedmannian world, respectively. The moments of time $\rm
m$, $\rm eq$, $\rm E$, and $\rm rec$, mark the end of the
inflationary Big Bang stage, the onsets of DM and DE eras, and
recombination, respectively.}%
\label{Fig-1}
\end{figure}

The problem of determining geometrical properties of the early
universe was successfully solved at the turn of the 21st century
in the Cosmological Standard Model (CSM) which describes the
entire set of experimental and observational data in the energy
range $10^{-3} - 10^{12}$ eV. The CSM made it possible to restore
the initial state of the universe by a direct extrapolation to the
past with a mere assumption that General Relativity (GR) holds up
to the GUT scale ($\sim 10^{25}$ eV). Further extrapolation to
higher energies is rather problematic because of the inflationary
Big Bang stage. At the latter stage the Hubble radius outgrows the
light horizon of the past where most information about
pre-inflationary geometry of the flow is stored (see
Fig.~\ref{Fig-1},~\cite{lukash2010}). Going to the past deviations
from the quasi-Friedmannian model increase during inflation.
Hence, the cosmological flow structure at the onset of inflation
could be well too different from the Friedmannian one and have a
different symmetry and topology.

By virtue of the CSM the problem of generating the initial
expanding flow (the cosmogenesis problem) has come into strict
scientific domain, because energies do not cross the Planck scale.
Besides, since ultra-high energies and curvature occur in the
gravitating system during only a short period of time, in studying
models with collapse turning into expansion it is sufficient to
use only local conservation laws which may be written in the
general geometric form of the Bianchi identities. To do so, any
modification of gravity or quantum-gravitational corrections are
included in the right-hand side and ascribed to an {\it effective}
energy--momentum tensor, thus, containing both material and, in
part, spacetime degrees of freedom. This approach allows us to
keep the notion of mean metric space--time (regardless of density
and curvature values) and stay in the class of geometries with
{\it integrable} singularities, which makes it possible to
construct geodesically complete maps of black/white holes and
understand how the $T$-region of a black hole is gravitationally
transformed into anticollapse of a newly generated matter in the
white hole (cosmological expansion).

We review below lessons taught by the extrapolation, determine
initial conditions in the early universe, discuss the physical
nature of the multi-sheet universe and present new models of
cosmological flow generation in the framework of the cosmogenesis
paradigm we have proposed (see~\cite{ls}).

\section{Lessons of extrapolation}

\begin{figure}[t!]
\begin{center}
{\includegraphics{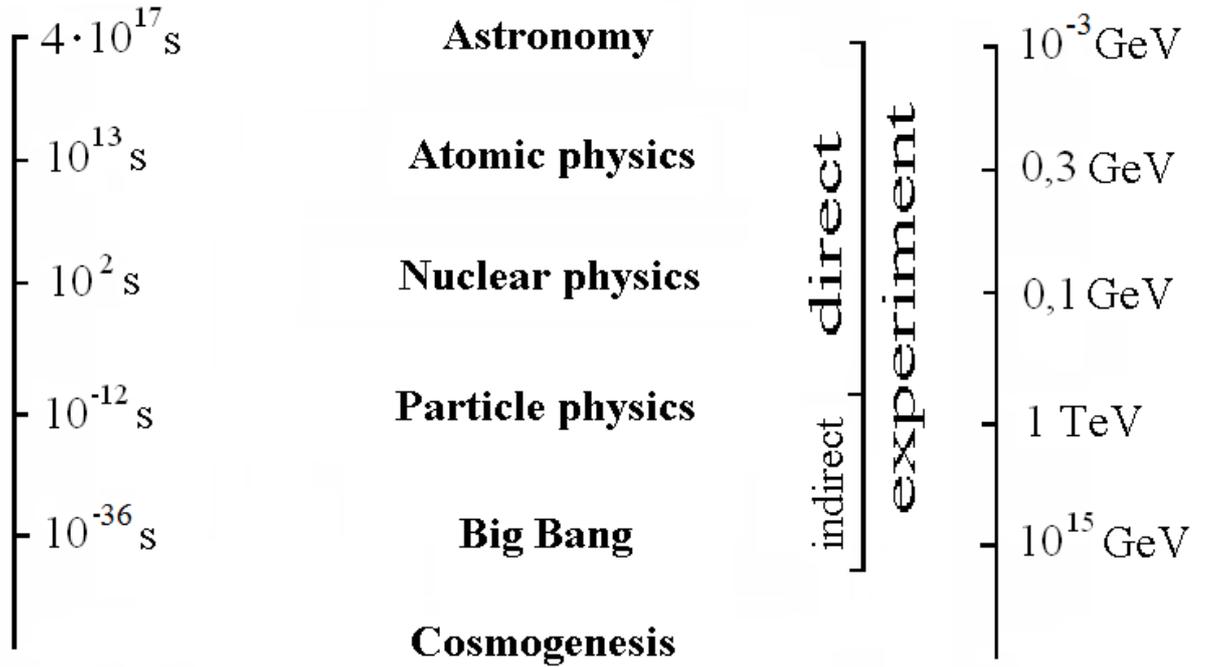}}
\end{center}
\caption{Experimental basis of the early universe. The curvature
radius (in seconds) of the universe is on the left vertical axis,
characteristic energies are on the right axis.}%
\label{Fig-2}
\end{figure}

The cornerstone of the CSM is a vast observational and
experimental basis spanning 15 orders of magnitude in energy: from
the current cosmological density ($\sim 10^{-3}$ eV) to the
electroweak scale ($\sim 10^{12}$ eV) studied at the Large Hadron
Collider and corresponding to the age of the universe of a few
picoseconds (see Fig.~\ref{Fig-2}). In the present time this is
established scientific knowledge rather than extrapolation. The
extrapolation begins at higher energies and spans another 13
orders of magnitude up to the GUT scale.

The most important result of the contemporary research is our
knowledge of the geometry of the early universe, which is
equivalent to the structure of metric and energy--momentum tensors
in GR. The empirically developed model has a small parameter --
the amplitude of cosmological metric inhomogeneities ($\sim
10^{-5}$), which allows us to exploit perturbation theory
techniques. In the zeroth order we deal with the spatially flat
Friedmann model described by a single function of time -- the
scale factor $a(t)$ defined by matter contents. In the first order
the tensors' structure is more complicated and expressed through
three irreducible forms~\cite{Lifshits}: scalar (density
perturbations), tensor (gravitational waves) and vector (related
to magnetic fields, for example) ones. Each of them is
characterized by its power spectrum, $S(k)$, $T(k)$ and $V(k)$,
where $k$ is the wave number (inverse scale of perturbation). With
spatial phase being random, the second order and higher ones do
not contain new free functions.

Therefore, we conclude that the initial cosmological matter flow
is fully deterministic and possesses a laminar quasi-Hubble
structure (weakly inhomogeneous, or quasi-Friedmannian universe).
Provided that the initial conditions and matter contents are set,
the world develops into the rich palette of physical processes and
phenomena we currently observe. At the moment we know two first
functions of the four mentioned above in the range available to
observational cosmology. If the up-and-running PLANCK experiment
succeeds the power spectrum of gravitational waves will be
revealed as well. Detection of the vector mode is beyond our
experimental capability so far.

The main goal of the cosmogenesis problem is to explain the
starting properties of the cosmological flow. Its physical
formulation stems from the lessons of extrapolation in the
past~\cite{lukash2010}. Seven of them we review below.

 1) {\it The universe is large}. This fact can be explained by the
 short-period inflationary Big Bang stage that precedes the
 radiation-dominated expansion era.

The current Hubble radius (4-curvature radius) is $H^{-1}_0\simeq
4.3$~Gpc, which, on the time scale, lies 60 orders of magnitude
away from the Planck value. According to the CSM, during this
period the scale factor could grow mere 30 orders as confirmed by
the Friedmann equations describing the main order of perturbation
theory:
\begin{equation}\label{1}
H\equiv\frac{\dot a}{a}=H_0\sqrt{\frac{10^{-4}}{a^4}+
\frac{0.3}{a^3}+0.7}\;\; \rightarrow\;\frac{H_0}{100a^2}\,,
\end{equation}
\[
\gamma\equiv-\frac{\dot H}{H^2}=\frac{2\cdot 10^{-4}+0.4\cdot
a}{10^{-4}+0.3\cdot a+0.7\cdot a^4}\;\in\;\left(0.4\,,\,2\right)
\]
(three terms under the square root correspond to radiation,
non-relativistic matter and dark energy; the scale factor is
measured in units of its modern value). Extrapolating to the past
we obtain the universe dominated by radiation. Its size at early
times is at submillimeter scale, which is extremely big being 30
orders of magnitude greater than the Planck scale. In order to
explain this size one needs to introduce a preceding inflationary
stage with $\gamma< 1$ and no less than 70 Hubble epochs ($\sim
30\ln 10$).

 2) {\it Causality principle} is independent evidence of the inflationary Big Bang stage.

As follows from eqs.~(\ref{1}), in the radiation-dominated era
galactic scales appear to reside in causally disconnected zone
(see Fig. 1). They could enter this zone from a causally connected
region if there had existed a short inflationary stage.

 3) {\it Small tensor mode} (indicating the inflationary Big Bang as well) and a Gaussian field of density perturbations.

While the zeroth order of perturbation theory is described by the
Friedmann equations the first order consists of oscillators (see
Appendix). The modes $S$ and $T$ evolve as massless scalar fields
$q=(q_S, q_T)$ under the action of the external gravitational
field of the non-stationary Hubble flow, which leads to parametric
amplification of the $q$ fields in the course of cosmological
expansion~\cite{lukash1980, lukash2010}. Under quite general
assumptions on the expansion rate the equations governing the
behavior of the elementary oscillators yield a general solution
with excitation amplitudes depending on initial conditions. For
oscillators initially occupying the ground state the power spectra
of the generated perturbations are the following:
\begin{equation}\label{2}
T(k)\simeq\frac{H^2}{M_P^2}\;<\; 10^{13}\;\mbox{GeV}\;,\qquad
\frac{T}{S}\simeq 4\gamma\;<\; 0.1\;,
\end{equation}
where $\langle q^2_{S,T}\rangle=\int (S,T)\frac{dk}{k}$\,. The
brackets $\langle \ldots \rangle$ stand for averaging over the
state, $M_P\equiv G^{-1/2}\simeq 10^{19}$~GeV -- the Planck mass.
As one can see the theory does not discriminate between the tensor
and scalar modes while their ratio depends on the value $\gamma$
in the parametric amplification era. Non-equalities in (\ref{2})
reflect current observational bounds on cosmological gravitational
waves. The second inequality indicates that $\gamma$ in the early
universe was less than one, which is indirect evidence of the
inflationary start of the Hubble flow. Hard evidence of primordial
inflation will become available after the tensor mode is detected
in observations of cosmic microwave background and the predicted
relation between the $T$-spectrum index and the tensor-scalar
amplitude ratio is confirmed ($n_T\equiv d\ln T/d\ln
k\simeq-2\gamma \simeq -0.5\cdot T/S$).

It is worth pointing out that the latter conclusion is based on
the hypothesis that the early Hubble flow was ideal, which implies
the vacuum initial condition for $q$-fields. The assumption is
justified by the fact that, first, the observed random spatial
distribution of large-scale density perturbations is {\it
Gaussian} (a property of quantum fluctuations linearly transferred
to the field of inhomogeneities) and, second, the time phase of
acoustic oscillations corresponds to the {\it growing} adiabatic
evolution mode (a property of the parametric amplification
effect).

 4) {\it Dark matter presence.}

Nonlinear halos hosting galaxies consist of non-relativistic
particles of dark matter (DM) that neither interact with baryons
nor with radiation. The nature of DM particles is unknown, but
there are observational indications that the origin of DM is
related to the baryon asymmetry of the universe. Here are two of
them: cosmological mass densities of DM and baryons are close to
each other (the ratio is 5) and scales of their spatial
large-scale distributions coincide (the cosmological horizon at
the moment of equal densities of relativistic and non-relativistic
components is identical to the sound horizon at the era of
hydrogen recombination). If we take into account that the density
ratio for the two non-relativistic components does not change with
time we have to conclude that the reasons that led to generation
of DM and to baryon asymmetry are interconnected. Both DM
particles and extra baryons may have emerged in non-equilibrium
processes of particle transformation in high-temperature radiation
plasma of the Hubble flow. If this is the case, their origin has
nothing to do with pre-inflationary history of the Big Bang.

 5) {\it Evidence of dark energy}.

The matter forming the structure of the universe is tracked by
gravitational potential gradient in dynamical observations of
galaxies and gas and by gravitational lensing. Its amount does not
exceed $30\%$ of the critical density. The rest $70\%$ reside in
homogeneously distributed medium that does not interact with light
and baryons. This is the so-called dark energy (DE) with negative
effective pressure whose absolute value is comparable to the DE
density. We have probably encountered a relic ultra-weak field
which had remained ``frozen'' at the radiation- and
matter-dominated eras and then started slowly rolling under the
action of its own gravity 3.5 billion years ago. If it is true we
are witnessing relaxation of the massive field, which opens a new
look on the history of the Hubble flow.

 6) {\it History of the universe evolution}.

One can see that the history of evolution includes periods of
accelerated ($\gamma < 1$) as well as decelerated ($\gamma > 1$)
expansion. The former include inflationary stages of the Big Bang
and DE and the latter -- the radiation- and matter-dominated eras.
We know, however, that small perturbations decay if $\gamma < 1$
and grow otherwise. Thus, it happens that in the history of the
universe there were stages of {\it forming} (restoring) and {\it
decaying} Hubble flow (in the latter case the structure is
formed). This feature reveals a dual nature of long-range gravity
capable of creating highly ordered configurations from quite
general initial distributions and types of matter. Those are
anticollapse, or inflation (formation of the ideal Hubble flow)
and its antipode collapse (formation of gravitationally bounded
halos and black holes). Therefore, we can look on the dynamical
history of the flow as a 14-billion history of massive scalar
fields relaxing to their minimal-energy states. Here comes the
seventh and last lesson of extrapolation of the CSM to the
pre-inflationary universe. How can the conditions necessary for
the expanding matter flow to emerge and inflate into the observed
Hubble flow, be created?

\section{Cosmogenesis conditions}

As a matter of fact, any solution of the cosmogenesis problem must
answer three questions:
\begin{itemize}
\item{How do the high densities emerge?} \item{What triggers the
expansion?} \item{What is the origin of the cosmological
symmetry?}
\end{itemize}
Inflation does not address these questions. In its different
models (e.g.~\cite{guth, linde}) new physical fields are
introduced in an ultra-dense state from the very beginning. The
birth of the universe from ``nothing'' \cite{zeldovich, gibbons}
also involves the notion of a highly dense ``false'' vacuum, so do
models with modified gravity~\cite{starobinsky}. It is true that
in the so-called bouncing models, which have been developed for
more than 40 years, the problem of initial conditions does not
arise at all (thanks to modifications of equation-of-state), but
again the Friedmannian symmetry is postulated.

The fundamental scientific principle that states that any physical
solution describing nature must contain only such observable
quantities that remain finite, appears to be of great use in the
cosmogenesis problem. Indeed, if we consider realistic models of
black/white holes with {\it smoothed} metric singularities this
allows us to constrain the tidal forces (despite a possible
divergence of some curvature components) and construct a
geodesically complete metric space--time on the basis of dynamical
solutions resulting from the energy--momentum
conservation~\cite{ls}. Here the singularity emerging around the
collapsed object is surrounded by an effective matter. We model
the latter in a wide class of equations-of-state. Now the radial
geodesics pass to the $T$-region of the white hole rather than end
in the singularity. From this point we arrive to a hypothesis that
any black hole that has originated from the collapse of an
astrophysical object may give birth to a new (daughter, or
astrogenic) universe.

This conjecture easily solves all of the three above-mentioned
problems of cosmogenesis:

\begin{itemize}
\item{The ultra-high curvature and density on the initial stage of
cosmological evolution are achieved as a consequence of
superstrong and highly variable gravitational fields that exist
inside the black/white hole and generate a matter the daughter
universe consists of.} \item{The initial push to the expansion of
the generated matter (the Big Bang) is provided by the $T$-region
of the white hole. The initial cosmological impetus is, hence, of
pure gravitational nature and one of the manifestations of
gravitational (tidal) instability.} \item{The $T$-region symmetry
of the black hole outside the maternal matter of the collapsing
object is that of an anisotropic cosmology. It is transferred to
the white-hole $T$-region and can be made isotropic by the known
inflationary mechanisms.}
\end{itemize}

\section{Black/white holes with integrable singularity}

The above-mentioned principle applied to spherically symmetric
metrics of general type practically implies finiteness of the real
functions $N$ and $\Phi$\, in \,$\mathbb{R}^2\in (r,t)$:
\begin{equation}
\label{general-metrics}
ds^{2}=N^2(1+2\Phi)\,dt^{\,2}-\frac{dr^{\,2}}{1+2\Phi}-
r^2d\Omega\,,
\end{equation}
where $r$ and $t$ are, respectively, radial and time coordinate in
$R$-regions of the space-time ($\Phi >-1/2$) and, vice versa, time
and radial coordinate of the same solution in $T$-regions ($\Phi
<-1/2$, see~\cite{novikov}) while $d\Omega$ is the squared line
element on the surface of 2-dimensional sphere.

The GR equations yield:
\begin{equation}
\label{rm} \Phi = -\frac{G m}{r}\,,
\end{equation}
where the finite {\it mass function}
\begin{equation}
\label{m} m=m\!\left(r,t\right)= 4\pi\!\int_{0} T_t^t r^2 dr
\end{equation}
vanishes on the inversion line $r=0$ thanks to the finiteness
condition applied on the potential $\Phi$. $T_t^{t}$ is the
$tt$-component of the energy--momentum tensor which can be written
as $T_{\mu}^{\nu}= diag(-p,\, \varepsilon,-p_{\perp},-p_{\perp})$,
provided spatial flows in the $T$-region are absent. Integrability
of the function $T_t^t r^2$ at zero (which also follows from the
finiteness condition) leads us to the definition of {\it
integrable} singularity $r=0$ surrounded by an effective
matter\footnote{We suggest that this matter may be generated by
strong highly variable gravitational field (as a result of
quantum-gravitational processes) beyond the collapsed object in
the $T$-regions of the black and white holes. Then the symmetry of
the complete solution respects the global Killing $t$-vector
already present in the original Schwarzschild vacuum metrics, and
all physical quantities under consideration are functions of $r$
alone (we set $r>0$ in the maternal black hole and $r<0$ after the
continuation of the metrics across the line $r=0$).}. We give
below two examples of the models in which energy density is
generated through variations of the transversal pressure
$p_{\perp}$ changing in a triggered way at certain moments of time
$r$. These models have finite tidal forces along the radial
geodesics, and world lines of test particles are continued from
the $T$-region of the black hole to that of the white hole. In
other words, the tidal gravitational interaction in the vicinity
of the integrable singularity undergoes an oscillation in time
which connects the interiors of both holes. This phenomenon can be
referred to as a collapse {\it inversion}.

\section{Astrogenic universes}\label{universes}

The matter in the $T$-regions of the vacuum solutions can be
generated through time variations of the function $p_\perp(r)$,
e.g. through discontinuities of first kind, since
equations-of-motion do not contain its derivatives (energy density
is pumped from the gravitational field and the metrics is
consistently reconfigured to satisfy GR). For simplicity the
longitudinal pressure is conveniently chosen to be vacuum-like
($p=-\varepsilon$). Hence, $N=1$ everywhere in $\mathbb{R}^2$ and
the matter is at rest in the reference frame
(\ref{general-metrics}), the energy density being found from the
Bianchi identities:
\begin{equation}
\label{tB} \frac{d\left(\varepsilon r^2\right)}{rdr}=-2p_\perp\,.
\end{equation}

\begin{figure}[t!]
\begin{center}
{\includegraphics{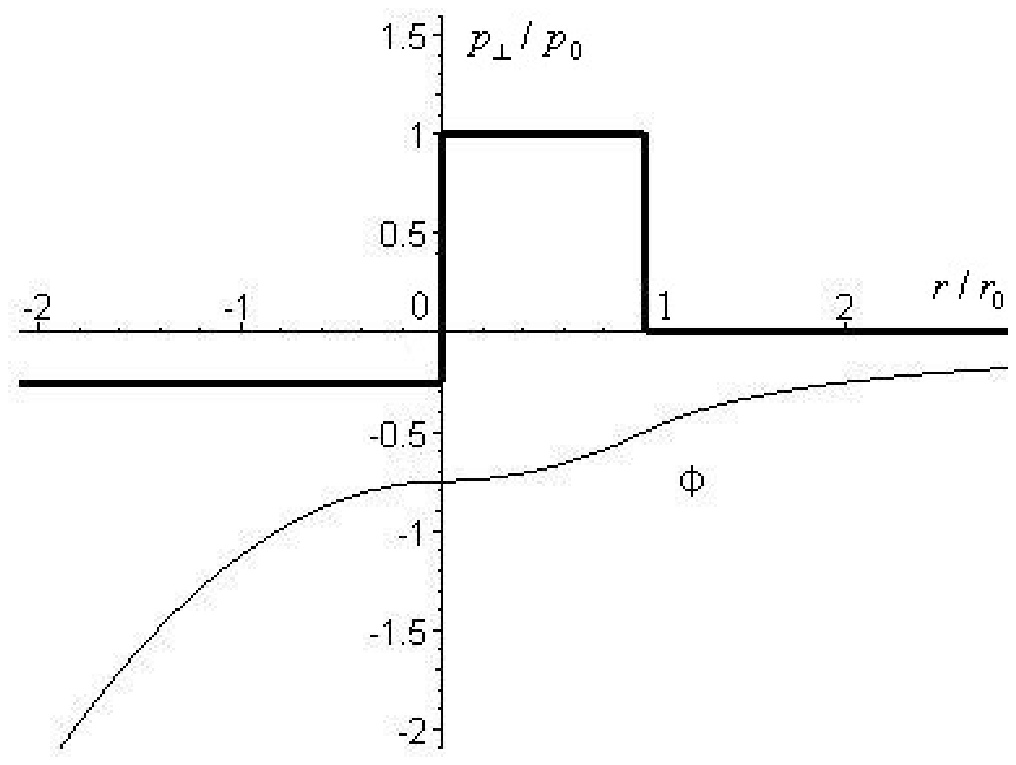}}
\end{center}
\caption{An asymmetric profile of the transversal pressure (in
bold) inverting collapse into cosmological expansion that is
asymptotically de-Sitter one. The thin line shows evolution of
gravitational potential. This plot corresponds to the model in
which the matter entirely fills $T$-region of the black hole
($r_{0}=2GM$). Besides, as an example, we set $p_{1}/p_{0}=0,5$
(see Sec.~\ref{universes}).}%
\label{pressure-A}
\end{figure}

\begin{figure}[t!]
\begin{center}
{\includegraphics{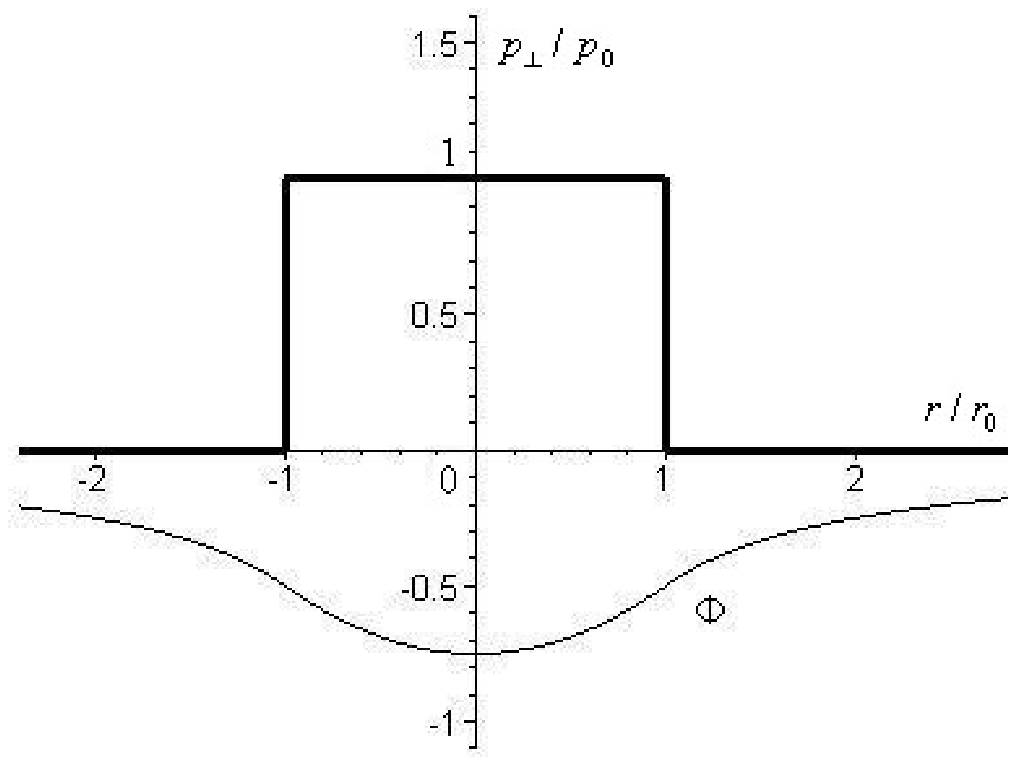}}
\end{center}
\caption{A symmetric profile of the transversal pressure (in bold)
turning the black hole into a white hole of the same mass. The
thin line shows the evolution of gravitational potential. The
matter entirely fills $T$-regions of both black and white holes ($r_{0}=2GM$).}%
\label{pressure-B}
\end{figure}

Let us consider two toy examples of the $p_\perp$-function
behavior (Figs.~\ref{pressure-A},~\ref{pressure-B}):
\begin{itemize}
\item[(A)]{an asymmetric step,\,
$p_\perp^{\left(A\right)}=p_0\cdot\theta\left(rr_0-r^{2}\right)-
p_1\cdot\theta\left(-r\right),$}
\item[(B)]{a symmetric step,\, $p_\perp^{\left(B\right)}=
p_0\cdot\theta\left(r_0^2-r^2\right),$}
\end{itemize}
where $r_0\le 2GM$ and $p_1$\, are positive real constants,
$M\equiv 8\pi r_0^3p_0/3\,$ -- the black-hole mass. Integrating
eq. (\ref{tB}) with initial condition $\varepsilon (r\ge r_0)=0$
yields the following functions $\varepsilon(r)$:
\begin{equation}\label{ep}
\varepsilon^{\left(A\right)}=-p_\perp^{\left(A\right)}+p_0\frac{r_0^2}{r^2}\cdot\theta\left(r_0-r\right),
\qquad
\varepsilon^{\left(B\right)}=p_\perp^{\left(B\right)}\cdot\left(\frac{r_0^2}{r^2}-1\right).
\end{equation}
Thus, $(A)$ is a model of the astrogenic universe
($\varepsilon^{\left(A\right)}\rightarrow p_1$ as
$r\rightarrow-\infty$) while $(B)$ is that of oscillating
(eternal) black/white hole. The potential $\Phi(r)$ is of the
$C^1$ class (see (\ref{rm}, \ref{m}),
Figs.~\ref{pressure-A},~\ref{pressure-B}).

Let us consider $(B)$ in two limits. As $r_0\rightarrow 0$ we
obtain a black/white hole maximally extended onto the empty space
with a delta-like source localized at $r=0$:
\begin{equation}
\label{d} \varepsilon=2p_\perp=M\,\frac{\delta\left(r\right)}{2\pi
r^2}\,,
\end{equation}
where $\delta(r)=\theta'(r)$ is one-dimensional delta-function.

In the limit $r_0=2GM$ we obtain a stationary black/white hole
with an oscillating matter flow in the $T$-region:
\begin{equation}\label{od} r=-2GM\sin\left(H\tau\right)\,,\qquad \varepsilon
=\frac{3H^2}{8\pi G}\,\cot^2\!\left(H\tau\right)\,,
\end{equation}
\[
ds^2=d\tau^2-\frac 12\left(\cos^2\!\left(H\tau\right)dt^2 +
\frac{\sin^2\!\left(H\tau\right)}{H^2}d\Omega\right),
\]
where $H^{-1}\equiv 2\sqrt{2}GM$ and $\tau$ are the oscillation
frequency and proper time of the flow, respectively.

\begin{figure}[t!]
\begin{center}
{\includegraphics{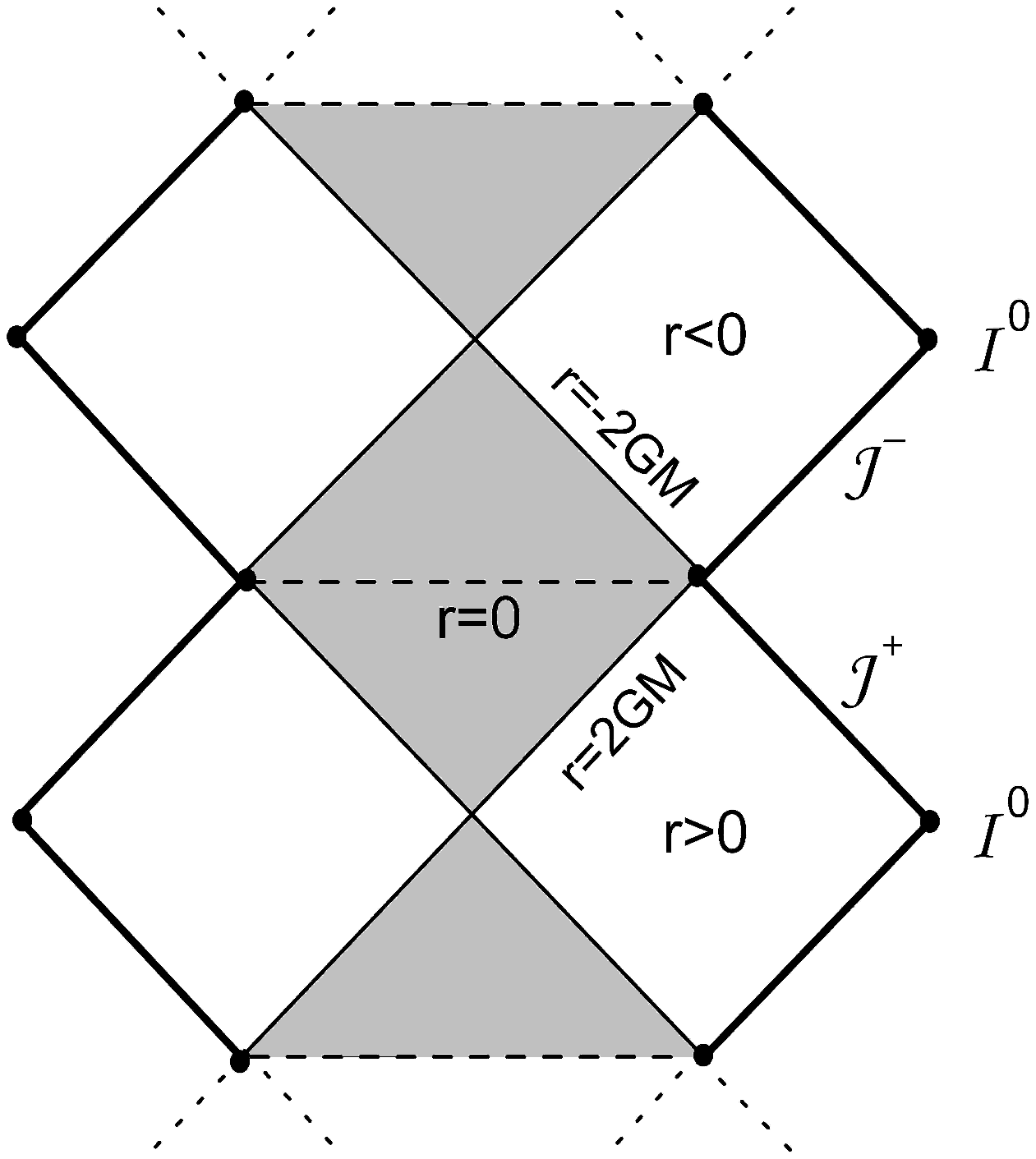}}
\end{center}
\caption{Penrose diagram of the pulsating flow with the symmetric
function $p_{\perp}(r)$ (see Fig.~\ref{pressure-B}). The matter
occupies the shaded region.}%
\label{penrose-B}
\end{figure}

Fig.~\ref{penrose-B} shows the Penrose diagram of this pulsating
matter flow, spatially homogenous and anisotropic. Phase
transitions in the matter on the stage of its expansion may, in
principle, cause inflation and isotropize the flow to the
Friedmannian symmetry in an arbitrary large volume.

\begin{figure}[t!]
\begin{center}
{\includegraphics{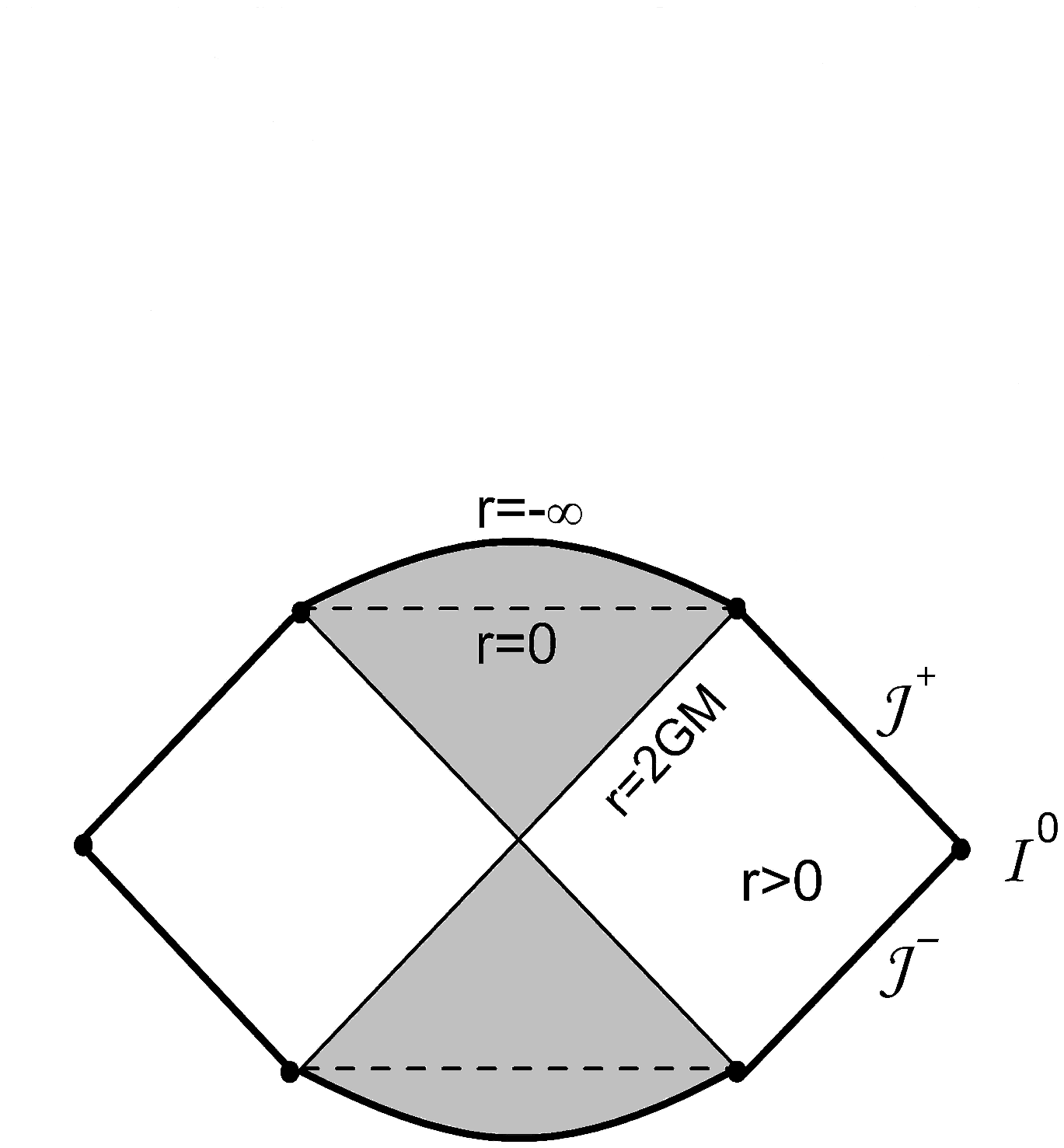}}
\end{center}
\caption{Penrose diagram of an astrogenic universe with the
asymmetric function $p_{\perp}(r)$ (see Fig.~\ref{pressure-A}).}%
\label{penrose-A}
\end{figure}

The simplest realization of this scenario is exemplified by the
case $(A)$. Indeed, if $\tau\ge 0$ eq. (\ref{ep}) yields the
solution asymptotically approaching de-Sitter (see
Fig.~\ref{penrose-A}):
\begin{equation}
\label{au} r=-\frac{\sinh(H_1\tau)}{\sqrt{2}H_1}\,,\qquad
\varepsilon =\frac{3H^2_1}{8\pi G}\,\coth^2\!\left(H_1\tau\right),
\end{equation}
\[
ds^2=d\tau^2-\frac 12\left(\coth^2\!\left(H_1\tau\right)dt^2+
\frac{\sinh^2\!\left(H_1\tau\right)}{H_1^2}d\Omega\right),
\]
where the constant $H_1= (8\pi Gp_1/3)^{1/2}$ acquires any value
independent of the external mass of the black hole. The presented
toy model of astrogenic universe can be further elaborated by
introducing massive scalar fields, radiation and the other
ingredients of the contemporary CSM.

\section{Conclusions}

Extrapolation of the CSM to the past indicates that the initial
Hubble matter flow expands from ultra-high curvature and density.
In the models of black/white holes with integrable singularities
the cosmological flows can emerge in the expanding white-hole
$T$-regions lying in the absolute future with respect to the
$T$-region of the maternal black hole. In the framework of this
paradigm we introduce the notion of {\it astrogenic cosmology},
which is a cosmology originated from inversion of the collapse of
some astrophysical compact system to expansion of the effective
matter flow outside the body of the collapsed object. Figuratively
speaking, black holes in these models are being lighters setting
the new worlds on fire.

Multi-sheet universes with intricate topology predicted and
discussed by A.D.~Sakharov could have been realized through the
collapse of compact systems on their final stages of evolution in
a maternal universe. As mentioned above, a universe born as a
result of this collapse needs to be isotropized (if we believe
that this universe is similar to ours). The reason for that is a
non-Friedmannian, though cosmological, symmetry of the interior of
the white hole, namely, the cylindrical symmetry
$\mathbb{R}\times\mathbb{S}^{2}$ of the Kantowski-Sachs model.
Hence, residual cylindrical anisotropy in the present-day data
would indicate the astrogenic mechanism of the beginning of our
universe. Although some authors claim to have discovered a global
anisotropy~(see~\cite{komberg}, for example), the current
precision of cosmological observations is insufficient to state
that, and future observations should clarify the situation.

\section{Acknowledgements}
The authors are grateful to the organizers of the Sakharov Session
for the opportunity to present the talk and to P.\,B.~Ivanov for
critical reading of the manuscript. This work was partly supported
by the Russian Foundation for Basic Research (project codes
11-02-12168-OFI-M-2011, 11-02-00244) and by the grant of Ministry
of Science and Education of the Russian Federation (no.
16.740.11.0460). VNS also thanks the Dynasty Foundation for
financial support.

\newpage


\appendix

\section*{\appendixname}
Recall~\cite{lukash2010} that \,$q_S=\delta a/a+Hv$\, where\, $v$
are perturbations of the comoving scale factor and velocity
potential of the matter, respectively. $q_T=(q_\lambda)$ are
amplitudes of gravitational waves with polarizations
$\lambda=\oplus, \otimes$. The conformal fields $\tilde
q={\tilde\alpha} q/\sqrt{8\pi G}$ obey the equations of classical
harmonic oscillators with variable frequencies:
\begin{equation}
\label{ur} {\tilde
q}^{\,\prime\prime}+\left(\omega^2-U\right){\tilde q}=0\,,
\end{equation}
where the prime stands for derivative with respect to the
conformal time $\eta=\int dt/a$,
\[
U\equiv\frac{{\tilde\alpha}^{\prime\prime}}{\tilde\alpha\;}\,,\qquad
U_T=\left(2-\gamma\right)\!a^2H^2\,,\qquad {\tilde\alpha}_S=
\frac{a\sqrt{2\gamma}}{\beta}\,, \qquad {\tilde\alpha}_T=a\,,
\]
$\omega=\beta k,\; \beta_S$ -- is the sound speed in the
speed-of-light units, $\beta_T=1$. In the case when more than one
medium is present the right-hand side of the $S$-oscillator
equations acquires an additional term describing the action of
isocurvature perturbations.

The dependence of the effective frequency ($\omega^2-U$) on time
causes parametric amplification of the elementary oscillators in
the course of the universe evolution. Assuming the vacuum initial
state in the wave zone ($\omega^2> |U|$) and taking into account
that the latter then turns into the parametric one ($|U|
>\omega^2$) we obtain the solution of (\ref{ur}) in the form:
\begin{equation}
\label{sol} \frac{\exp{\left(-i\int\omega
d\eta\right)}}{{\tilde\alpha}\sqrt{2\omega}}\, \rightarrow\,
\frac{{\mathfrak c}-i}{C\sqrt{2k}}\, \rightarrow\,
\frac{M_P\sqrt\pi}{2k^{3/2}}\,q_k\,,
\end{equation}
where $C$ is a junction constant in the region
$|U|\simeq\omega$\,,\, the function ${\mathfrak c}=-k
C^2\!\int{\tilde\alpha}^{-2}d\eta\rightarrow const$ converges at
the upper limit if $\gamma< 3$. The ``frozen'' fields
$q_k=const(\eta)$ correspond to the growing mode of the general
solution. Their phases are random while their absolute values give
the spectral amplitudes $S=|q_{kS}|^2$ и
$T=|q_{k\oplus}|^2+|q_{k\otimes}|^2$. When $\beta=1$ and
$\gamma\simeq const$ eqs. (\ref{ur}) are identical for either mode
and $T/S=2{\tilde\alpha}_S^2/{\tilde\alpha}_T^2=4\gamma$. If
$\gamma < 1$ we obtain the $T$-spectrum (\ref{2}) within a
multiplicative factor of the order of unity.

\newpage

\end{document}